# On the Analogy of Gauge Theory of Plasticity and Economics

A. V. Samokish[1] and V. E. Egorushkin[2]

[1] *National Research University Higher School of Economics, Moscow, 101000, Russia*

[2] *Institute of Strength Physics and Materials Science, Siberian Branch, Russian Academy of Sciences, Tomsk, 634055 Russia*

**Annotation**

We demonstrated the analogy between Economics and Gauge Theory of Plasticity and used it to describe the relationship between money supply and inflation at the economic market. The received equations of economical dynamics in phase space are similar to the plasticity equations and economic variables – choice, competition and profit correspond to the state of the market with inflation. We described the meaning of equations and the role of its variables in the stabilization mechanism of the market with inflation. The equation of market equilibrium including the Profit turnover, time changes of competition, capital and choice was discussed in detail.

1. Introduction

There are a lot of articles devoted to penetration of Physics and its methodology to the different branches of economics [1-7]. Electrodynamics [1, 2], thermodynamics [3-5] and gauge field theory have given strong input to economics and been awarded with Noble Prize [8]. The paper [1] discusses the commonality between electrodynamics and simple economic systems and the identity of economic and electrodynamic variables. The authors emphasize that both variables and laws in these disciplines are similar. The paper describes Phillips curve, recession, Black-Scholes formula, demand, and supply curves on the basis of formulas similar to Maxwell's equations. In [2], the problem of taking into account possible financial instability in the labor market model, reflecting empirical data, such as the model of Lottka-Voltera, is considered. Here the Maxwell's equations in fractional space are also applied. In [3] J. Rosser discusses the relationship between econophysics and entropy as the basis for economic phenomena. General equilibrium growth theory, business cycles, urban-regional economics, income and wealth distribution and financial market dynamics were takin into consideration. It is shown how the power law of distribution in econophysics reflects the interdependence of entropic and anti-entropic processes with economic dynamics, financial markets and basis cycles.

 In [4] is shown that there is a deep analogy between thermodynamic variables and parameters of economic markets. The concepts of energy and temperature were justified and introduced into an economic theory. Economic thermodynamics allows to gain an adequate estimate of economic information. Analogy of thermodynamics and economics is based on the concepts and laws present in both sciences including work, heat, temperature as price level, entropy as quality referred to the human capital. The results are applied to the economic growth as well [5]. Complete statement of the application of quantum field theory methods to the economic problems is given in [6]. Analysis of relationship between the disciplines highlighted the parallels between money income and wealth, entropy and economic exchange [7]. Extending this analogy, we emphasize its multifaceted nature, indicating that these disciplines share profound interconnections.

In either case, the initial equilibrium dynamic behavior of the particles and the economic market is violated, so that the equations corresponding to such equilibrium [9] cannot describe the new conditions, no matter how they are corrected. It implies that the new conditions cause the dynamic trajectories of particles and monetary streams in the phase space to become discontinuous [9, 10]. The phase space in the economy, just like in physics, is an abstract space with orthogonal coordinates formed by variables determining the equilibrium economic state. For different economic phenomena, these variables may vary as well [10,18]. In the case of real estate management, these may be market "participants", "objects", and "place of events" of the real estate market, while for the financial market they may be dimensionless values of price changes for goods, the number of goods and the velocity of money turnover. The emergence of strong inflation in the latter case can lead to a discontinuity of trajectories that is impossible to deal with via the introduction of "real" interest rate targeting. [19]. That is, the simple correction of the equilibrium equation will no lead to dealing with the discontinuity. It implies that within the previous equilibrium, it is impossible to eliminate structural changes in the market (inflation makes the phase space heterogeneous [19]). In such cases, mathematical restoration of the continuity of trajectories occurs by introducing curveted or isotopic spaces in which the movement of particles, money supply, etc. is determined by parallel transport [11]. That is, the problem and the formulation of the problem in economics and physics are similar. The situations described above in physics occur both in electrodynamics and in the theory of plasticity of materials as well. Moreover, while the phenomenon of electricity itself is still a mystery, the plasticity has a clear nature associated with defects of elasticity field in a solid.

The investigation of plasticity using gauge field theory have shown that the plasticity defects can be described with gauge fields [12-16], analogous to electrodynamics. Plasticity implies that a local violation of symmetry in Euclidean space $\{x_\mu, t\}$ of elastic displacements caused by certain defects lead to an additional phase in $(x_\mu, t)$ of these displacements. This phenomenon is similar to the one present in electron electrodynamics implying that $\frac{\partial b(x,t)}{\partial x_\mu} = E_\mu$ – internal curvature of the trajectory with discontinuity and $\frac{\partial^2 b(x,t)}{\partial x_\mu \partial x_\nu} - \frac{\partial^2 b(x,t)}{\partial x_\nu \partial x_\mu} \neq 0$ – internal surface curvature involved in parallel transport. Curvature $E_\mu$ – vector potential of the gauge field and derivatives

$$J_\mu = -\frac{1}{c}\frac{\partial E_\mu}{\partial t} - \frac{1}{c}V_\beta \frac{\partial E_\mu}{\partial x_\beta}$$

-plastic flow – defect flow density,

$$\alpha_\mu = \varepsilon_{\mu\nu k}\frac{\partial E_k}{\partial x_\nu}$$

–вихрь curvature vortex(linear defect density tensor) .c is the flow velocity, $\varepsilon_{\mu\nu k}$ is the Levi-Civita symbol. The physical meaning of $J_\mu$ - the force acting from the defects on the sources - flows of elastic displacements. This term $V_\beta \frac{\partial E_\mu}{\partial x_\beta}$ represents a vortex force, similar to the Lorentz force in electrodynamics.

In this case, plastic defects that disrupt the initial dynamic equilibrium for elastic displacements lead to a completely different, plastic equilibrium for $J_\mu$ and $\alpha_\mu$. The latter is due to the plastic driving force - the curvature of the trajectory of elastic displacements in the presence of discontinuities and their parallel transport.

Plastic processes are strictly local, unlike electrical processes that have long-range influence. Moreover, the sources of plasticity are discontinuities in the flows of other displacements, rather than magnitudes of charges, and all plastic processes are governed by the "rule of flows" [16]. These facts better align with economic market processes. In [13-16], solutions for local plastic flows and the patterns of their dynamics were found. An analogy with economics will allow us to transfer these results to economic processes of Choice, Competition, Profit, corresponding to the plastic variables $E_\mu, J_\mu, \alpha_\mu$, thereby making the "invisible hand of the market" a visible one.

## 2. Equations of Plastodynamics

Plastodynamics equations for dimensionless vectors $J_\mu$, $\alpha_\mu$ in the differential form follow [13-16]:

$$\frac{\partial J_\mu}{\partial x_\mu} = -\frac{1}{c}\frac{\partial \ln u}{\partial t} \quad (1)$$

- Continuity equation for a medium with defects:

$$\varepsilon_{\mu\nu k}\frac{\partial J_k}{\partial x_\nu} = -\frac{1}{c}\frac{\partial \alpha_\mu}{\partial t} \quad (2)$$

- Compatibility equation of $J_\mu$ and $\alpha_\mu$.

$$\frac{\partial \alpha_\mu}{\partial x_\mu} = 0 \quad (3)$$

- Vortex field condition with zero sources

$$\varepsilon_{\mu\nu k}\frac{\partial \alpha_k}{\partial x_\nu}E = \rho\frac{\partial J_\mu}{\partial t} + \sigma_\mu - E_\nu C_{\mu\nu} \quad (4)$$

- Constitutive plastic equilibrium equation. $\sigma_\mu = C_{\mu\nu}\frac{\partial \ln u}{\partial x_\nu}$ - elastic stress concentrators defined by Hooke law, $\rho$ – material density, $E$ – Young modulus, $C_{\mu\nu}$ - elastic moduli. Equation (4) represents the sum of all forces generated by possible plasticity mechanisms [16]. A detailed explanation of equations (1) - (4) is provided in [13-16].

## 3. Correspondence of Physical and Economic Variables

Match economic and physical variables from equations (1) – (4) in accordance with [1]:
1. Money density (M) – elastic displacements ($u_\mu$)
2. Money flow ($\frac{\partial \ln M}{\partial t}$) – flow of displacements ($\frac{\partial \ln u}{\partial t}$)
3. Capital ($\frac{\partial \ln M}{\partial x_\mu}$) – elastic stress concentrators ($C_{\mu v} \frac{\partial \ln u}{\partial x_v}$).
4. Choice flow ($Ch_\mu$) – plastic distortions ($E_\mu$)
5. Competition flow ($C_\mu$) – plastic flow ($J_\mu$)
6. Profit flow ($\Pi_\mu$) – defects density ($\alpha_M$)
7. Price index ($P_i$) – scalar potential (P)
8. Inflation ($\pi = \frac{\partial P_i / \partial t}{P_0}$) – work of outside forces to defects transport

## 4. Equations of Market Dynamics under Inflation

Similarly to (1) the first market equation for dimensionless variables takes the form:

$$\frac{\partial C_\mu}{\partial x_\mu} = -\frac{\partial \ln M}{\partial t} \quad (5)$$

Where $C_\mu$ – flow of competition, $\frac{\partial \ln M}{\partial t}$ – growth rate of money density (DM), i.e. the flow of DM. From the economic point of view, equation (5) implies that surplus (shortage) of liquidity generating the competition and stimulate activity in the market. Specifically, growth of the money density acts as a source for competition, while the increase in the competition flow hinders the DM growth.

The following equation, similar to (2), links the turnover (rotation) of the competition flow to the evolution of the profit flow:

$$\varepsilon_{\mu v k} \frac{\partial C_k}{\partial x_v} = -\frac{\partial \Pi_\mu}{\partial t} \quad (6)$$

Equation (6) demonstrates that the profit flow is one of the forms of competition and reflects a dual relationship between $C_\mu$ and $\Pi_\mu$. That is, neither $C_\mu$ nor $\Pi_\mu$ act as sources of each other. From the standpoint of competition, whose source is the increase in DM, it hinders the change in profit growth. From the perspective of profit, which is determined by the turnover (rotation) of the choice flow, the increase in profit flow hinders changes in the turnover of competition. That is, equation (6) indicates that the turnover of the competitive flow tends to reduce the profit flow. More precisely, when there is a tendency for profit to increase, competition tends to reduce profit (quite in the spirit of Adam Smith). When there is a tendency for profit to decrease, competition tends to increase it. Relationship (6) does not define the sources and mechanisms for the formation of $C_\mu$ and $\Pi_\mu$. However, it is one of the external regulators of market equilibrium relative to these sources with information, which can be arbitrarily large, but not destructive to these market relations.

The equation for the profit flow has the double meaning and takes the following form:

$$\frac{\partial \Pi_\mu}{\partial x_\mu} = 0 \quad (7)$$

On the one hand, (7) indicates the absence of profit sources among the money flows, so that the lines of profit flow come to an end at the market boundaries. On the other hand, the equation indicates that the overall change in profit is equal to zero. That is, if something increases profit, something else should decrease it. This could be the capital involved in profit creation and the choice flow (or labor) leading to its diminishing. The latter is reflected in the equation analogous to equation (4):

$$\varepsilon_{\mu\nu k} \frac{\partial \Pi_k}{\partial x_\nu} E = \rho \frac{\partial C_\mu}{\partial t} + K_\mu - Ch_\mu \quad (8)$$

The equation (8) connects the market profit turnover given by the left hand side of the equation with competition flow, capital: $K_\mu = \frac{C_{\mu\nu}}{E} \frac{\partial \ln M}{\partial x_\nu}$ and the choice flow $Ch_\mu = Ch_\nu * C_{\mu\nu}$. Equation (8), as well as other equations of dynamic equilibrium, implies that the sum of all forces created by possible market dynamics mechanisms is equal to zero. Each term in (8) corresponds to a specific mechanism related to the dynamics of capital, profit, competition and choice flows. Furthermore, (8) can imply that the choice flow contributes to the growth of competition and capital but suppresses profit creation.

The entire sequence of market processes obeys the *rule of flows*, similar to [16]. Expressing equations (4) - (8) in the integral form to understand it properly:

$$\oint_S C_\mu^{(i)} dS_l = -\frac{\partial \varphi}{\partial t} \quad (9)$$

$$\oint_l C_\mu^{(i)} dl = -\frac{\partial \theta_\mu}{\partial t} \quad (10)$$

$$\oint_S \Pi_\mu^{(i)} dS_\mu = 0 \quad (11)$$

$$\oint_l \Pi_\mu^{(i)} dl = \frac{\partial}{\partial t} \oint_{S_l} C_\mu^{(i)} dS_l + \oint_{S_l} K_\mu^{(i)} - \oint_{S_l} Ch_\mu dS_l \quad (12)$$

Where $S_l$ - the surface bounded by the contour L of the circulation $\Pi_\mu$ inside the market. $S$ – the surface enveloping the entire market volume, $\varphi = \int \ln m \, dv$, $\theta_\mu = \int \Pi_\mu \, dS_l$ – Berry phase [17] representing the total profit.

Equations (9) - (12) indicate that the sequence of market processes in the presence of inflation follows the *rule of flows*:

(9) – The flow of money within a given market area, hindering the flow of competition across the surface that bounds this area.

(10) – The flow of total profit (Berry phase) hindering the change in the economic driving force - the circulation of competition along the contour L within the market.

(11) – The flow of profit (internal curvature of the surface S) does not cross this surface meaning that the profit sources are absent. The origin of profit is not caused by the curvature of the trajectory but by its torsion - the internal curvature of the surface enveloped by the trajectory.

(12) – represents the sum of all flows passing through the surface $S_l$. Similar to Kirchhoff's law in electrodynamics, the sum of all economic flows through the boundary of $S_l$ is equal to zero. The difference in flows ($K_\mu - C_{h,\mu}$) is established by the lower limit of capitals' influence, after it has been spent to satisfy choices (self-interest). That is,

$$K_\mu{}^{min} = Ch_\mu = \frac{1}{2}\varepsilon_{\mu kl} \oint x_k \Pi_l dS_i \frac{C_\mu}{E} \quad (13)$$

The value of $K_\mu{}^{min}$ is determined by profit, relative market elasticity, and its size $x_k$. Capital $K_\mu$ starts to effectively operate in the market when $K_\mu > K_\mu{}^{min}$. At the lower limit, i.e. when $K_\mu = K_\mu{}^{min}$, the equation becomes the following:

$$\oint_l \Pi_\mu dl = \frac{\partial}{\partial t} \oint C_\mu dS_l \quad (14)$$

- The rate of change of the competition flow through surface $S_l$ maintains the circulation of profit around the contour enveloping this surface.

This is exactly the mechanism of the market's "invisible hand" - a self-sustaining state where the aggregate flow of non-stationary competition supports the circulation of profit. The area of the market where this occurs acts as a "resonator" for waves of competition and profit. To visualize this mechanism, we need to find solutions to equations (5)-(8). However, before solving and analyzing these equations, we should note that the applied analogy between economic, plastic, electrodynamic, and other phenomena has a profound meaning. It is based on the fact that these trajectories are distorted due to discontinuities in the trajectories of sources - charges, flows, elastic displacements, flows of money and others. Hence the linear movement of sources transitions into parallel transport in curvilinear coordinates. This general situation underlies the indicated analogy. In all cases, during such movement, an additional heterogeneous phase (Berry phase) is created for charges [17], the Burgers vector for dislocations [16], and the total profit $\Pi_\mu = \int \overline{\overline{\Pi}} d\overline{S_l}$, where $S_l$ – the surface of the active market. Also, $\frac{\partial \Pi_\mu}{\partial x_\mu} = Ch_\mu$ – the principal curvature of the money flow trajectory, choice flow: $\frac{\partial^2 \Pi_\mu}{\partial x_\mu \partial x_\nu} - \frac{\partial^2 \Pi_\mu}{\partial x_\nu \partial x_\mu} \neq 0$ – the internal curvature of the surface enveloped by the trajectory determining the system dynamics. Curvature $C_h$ is the vector potential of the gauge field. Its derivatives $C_\mu = -\frac{\partial Ch_\mu}{\partial t}$ and $\Pi_\mu = \varepsilon_{\mu\nu k} \frac{\partial Ch_k}{\partial x_\mu}$ represent the field strength and define the flows of competition and profit. From the definition of $C_\mu$ and the expression (6) it follows that $C_\mu$ is defined with precision up to the gradient of particular function, i.e.:

$$C_\mu + \frac{\partial Ch_\mu}{\partial t} = -\frac{\partial \varphi}{\partial x_\mu} \quad (15)$$

Where Φ – scalar potential (similarly to electrodynamics [1], plasticity [16]) that stands for the work of outside forces aimed at the. In economics, $\varphi = P$ that determines the price [1] and inflation levels: $\pi = \frac{\frac{\partial P_i}{\partial t}}{P_0}$. Taking this into account, equation (15) can be modified as following:

$$C_\mu = -\frac{\partial Ch_\mu}{\partial t} - \frac{\partial P}{\partial x_\mu} \quad (16)$$

The flow competition consists of two components: competition through the evolution of choice ($C_\mu^{(1)} = \frac{\partial Ch_\mu}{\partial t}$) and competition through price changes ($C_\mu^{(2)} = \frac{\partial P}{\partial x_\mu}$). Both $C_\mu^{(1)}$ and $C_\mu^{(2)}$ counteract the corresponding changes. Competition $C_\mu^{(1)}$ regulates the flow of choice in the market, while the price competition $C_\mu^{(2)}$ is driven by the actions of participants in market relations. The magnitudes are related to each other by a normalizing ratio (similar to electrodynamics and plasticity).

$$\frac{\partial Ch_\mu}{\partial x_\mu} + \frac{\partial P}{\partial t} = 0 \quad (17)$$

By differentiation and taking the gradient we obtain the relationship between $C_\mu^{(1)}$ and $C_\mu^{(2)}$:

$$\frac{\partial^2 C_\mu^{(1)}}{\partial x_\mu^2} = -\frac{\partial^2 C_\mu^{(2)}}{\partial t^2}; \quad \frac{\partial^2 C_\mu^{(2)}}{\partial x_\mu^2} = -\frac{\partial^2 C_\mu^{(1)}}{\partial t^2} \quad (18)$$

With an accelerated (decelerated) evolution, price and non-price competition suppress (support) each other. From (17) it follows that a change in the price level (inflation rate) act as the source of the choice flow. At the same time, the choice flow hinders the growth of the inflation rate. $C_\mu^{(1)}$ and $C_\mu^{(2)}$ are not connected to each other when prices change at a constant rate. With an accelerated (decelerated) inflation rate, $C_\mu^{(1)}$ and $C_\mu^2$ are connected by the relationship (18).

Therefore, the money supply does not directly control the inflation rate. It is connected to the flows of competition and profit, which regulate inflation. The relationships between these variables have been derived above, and their role in the market stabilization mechanism has been discussed. So, within the framework of the initial market equilibrium (e.g., the Fisher equation), changes in the money supply do not alter the equilibrium equation ("elastic market"). However, with the influence of choice flow, the market is deformed. This changes the economic dynamics, that is now determined by choice, competition, and profit according to the equations defined previously ("plastic" market). Profit acts as a "defect" of the initial market in the initial phase space, caused by the turnover of choice flow. Competition is determined by the choice flow and changes in the price level, i.e., inflation, that makes it possible to manage these flows. In further work, solutions to the economic equations (5) - (8) will be found.